# REGIONAL AGGLOMERATION IN PORTUGAL: A LINEAR ANALYSIS

**Vítor João Pereira Domingues Martinho**

Unidade de I&D do Instituto Politécnico de Viseu
Av. Cor. José Maria Vale de Andrade
Campus Politécnico
3504 - 510 Viseu
**(PORTUGAL)**
**e-mail:** vdmartinho@esav.ipv.pt

**ABSTRACT**

This work aims to study the Portuguese regional agglomeration process, using the linear form the New Economic Geography models that emphasize the importance of spatial factors (distance, costs of transport and communication) in explaining of the concentration of economic activity in certain locations. In a theoretical context, it is intended to explain the complementarily of clustering models, associated with the New Economic Geography, and polarization associated with the Keynesian tradition, describing the mechanisms by which these processes are based. As a summary conclusion, we can say which the agglomeration process shows some signs of concentration in Lisboa e Vale do Tejo (which is evidence of regional divergence in Portugal) and the productivity factor significantly improves the results that explain the regional clustering in Portugal (despite being ignored in the models of New Economic Geography).

**Keywords:** agglomeration; Portuguese regions; linear models.

**1. INTRODUCTION**

With this work, in a theoretical context, it is intended to explain the complementarily of clustering models, associated with the New Economic Geography, and polarization associated with the Keynesian tradition, describing the mechanisms by which these processes are based. It is pretended also studying the Portuguese regional agglomeration process, using the linear form the New Economic Geography models that emphasize the importance of factors in explaining the spatial concentration of economic activity in certain locations (1)(Martinho and Soukiazis, 2003).

One of the most recent publications of (2)Fujita, Krugman and Venables (Fujita et al., 2000) presents itself as a good contribution for the systematization of these developments, which can be summarized in a diagram like figure 1.

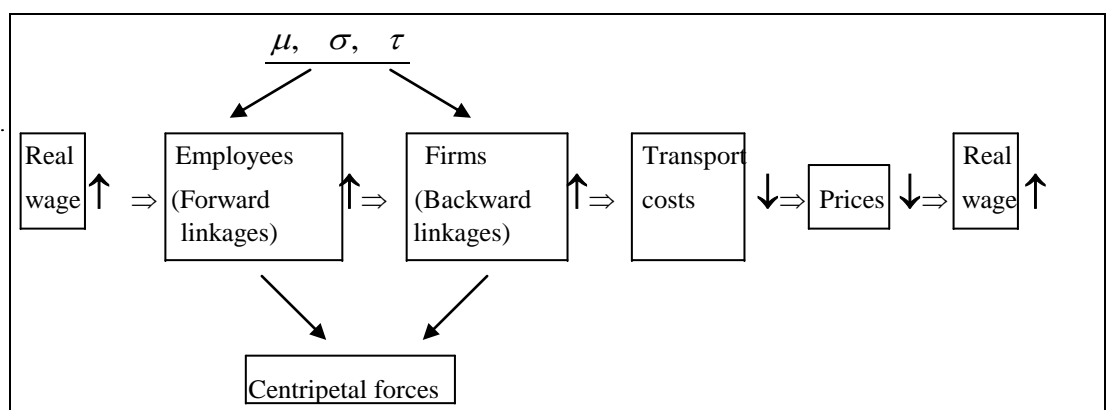

***Figure 1:*** *A mechanism that describes the process of agglomeration*

In Figure 2, is developed the mechanism that explains the process of polarization, based on the forces of demand, increasing economies of scale (in industry) and endogenous factors of production, among others (3)(Targetti et al., 1989). The relationship between productivity, better known as Verdoorn's law, makes the process of growth self-sustained with cumulative causes, circular and virtuous.

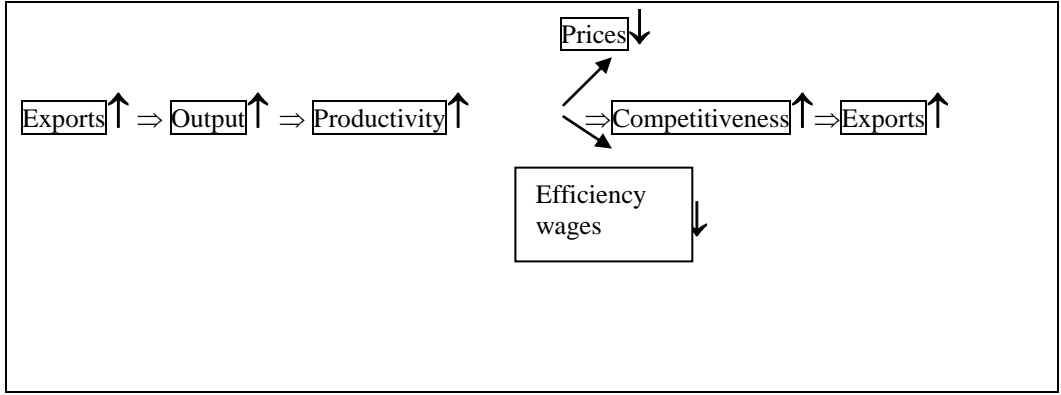

*Figure 2: A mechanism that describes the process of polarization*

## 2. DESCRIPTION OF THE MODEL, THE DATA AND THE ESTIMATES MADE

To analyze the goals set for this work, it was considered only the equation of real wages, from the equations of static equilibrium, in reduced form (equation (1)) and linear (equation (2)). The choice, only for this equation and in linear form, is due to the complexity found when working on the one hand, the equations in a nonlinear way, and secondly, a system of nonlinear equations.

$$\omega_r = \left[\sum_s Y_s T_{rs}^{1-\sigma} G_s^{\sigma-1}\right]^{1/\sigma} \left[\sum_s \lambda_s (w_s T_{sr})^{1-\sigma}\right]^{-\mu/(1-\sigma)}, \text{ equation of real wages reduced} \quad (1)$$

Linearizing the reduced equation (1), using logarithms, we obtain:

$$\log(\omega_r) = \frac{1}{\sigma}\log\left[\sum_s Y_s T_{rs}^{1-\sigma} G_s^{\sigma-1}\right] - \frac{\mu}{(1-\sigma)}\log\left[\sum_s \lambda_s (w_s T_{sr})^{1-\sigma}\right], \quad (2)$$

It should be noted that in the estimations made with the equations presented below, all variables were considered at national or regional level for 5 regions (NUTS II) of Portugal, and in time series of 8 years. All data for these variables were obtained from the regional database of Eurostat statistics (Eurostat Regio of Statistics 2000).

### 2.1. EQUATION LINEARIZED AND REDUCED OF THE REAL WAGES, WITH THE VARIABLES INDEPENDENT NATIONALLY AGGREGATED

Thus, the equation of real wages that will be estimated in its linear form, will be a function of the following explanatory variables:

$$\ln \omega_{rt} = f_0 + f_1 \ln Y_{pt} + f_2 \ln T_{rpt} + f_3 \ln G_{pt} + f_4 \ln \lambda_{pt} + f_5 \ln w_{pt} + f_6 \ln T_{prt} + f_7 \ln P_{rt}, \quad (3)$$

where:

- $\omega_{rt}$ is the real wage in region r (5 regions) for each of the manufacturing industries (9 industries);
- Ypt is the gross value added of each of the manufacturing industries at the national level;
- Gpt is the price index at the national level;
- $\lambda_{pt}$ is the number of workers in each industry, at national level;
- Wpt is the nominal wage for each of the industries at the national level;
- Trpt is the flow of goods from each of the regions to Portugal;
- Tprt is the flow of goods to each of the regions from Portugal;
- Prt is the regional productivity for each industry;

- p indicates Portugal and r refers to each of the regions.

The results obtained in the estimations of this equation are shown in Tables 1 and 2.

**Table 1:** Estimation of the equation of real wages with the independent variables aggregated at national level (without productivity), 1987-1994

$$\ln \omega_{rt} = f_0 + f_1 \ln Y_{pt} + f_2 \ln T_{rpt} + f_3 \ln G_{pt} + f_4 \ln \lambda_{pt} + f_5 \ln w_{pt} + f_6 \ln T_{prt}$$

| Variable | lnY$_{pt}$ | lnT$_{rpt}$ | lnG$_{pt}$ | ln $\lambda_{pt}$ | lnw$_{pt}$ | lnT$_{prt}$ | | |
|---|---|---|---|---|---|---|---|---|
| **Coefficient** | f$_1$ | f$_2$* | f$_3$* | f$_4$ | f$_5$* | f$_6$* | R$^2$ | DW |
| **LSDV** Coefficients T-stat. L. signif. | -0.038 (-0.970) (0.333) | 0.674 (4.227) (0.000) | -0.967 (-7.509) (0.000) | 0.025 (0.511) (0.610) | 0.937 (15.239) (0.000) | -0.594 (-3.787) (0.000) | 0.810 | 1.516 |
| Degrees of freedom | 290 | | | | | | | |
| Number of obervations | 302 | | | | | | | |
| Standard deviation | 0.146 **T.HAUSMAN - 416.930** | | | | | | | |

**(*) Coefficient statistically significant at 5%.**

**Table 2:** Estimation of the equation of real wages with the independent variables aggregated at national level (with productivity), 1987-1994

$$\ln \omega_{rt} = f_0 + f_1 \ln Y_{pt} + f_2 \ln T_{rpt} + f_3 \ln G_{pt} + f_4 \ln \lambda_{pt} + f_5 \ln w_{pt} + f_6 \ln T_{prt} + f_7 \ln P_{rt}$$

| Variable | lnY$_{pt}$ | lnT$_{rpt}$ | lnG$_{pt}$ | ln $\lambda_{pt}$ | lnw$_{pt}$ | lnT$_{prt}$ | lnP$_{rt}$ | | |
|---|---|---|---|---|---|---|---|---|---|
| **Coefficient** | f$_1$* | f$_2$* | f$_3$* | f$_4$* | f$_5$* | f$_6$* | f$_7$* | R$^2$ | DW |
| **LSDV** Coefficients T-stat. L. signif. | -0.259 (-7.064) (0.000) | 0.557 (4.422) (0.000) | -0.884 (-9.671) (0.000) | 0.256 (5.919) (0.000) | 0.883 (19.180) (0.000) | -0.493 (-3.996) (0.000) | 0.258 (10.443) (0.000) | 0.858 | 1.560 |
| Degrees of freedom | 289 | | | | | | | | |
| Number of obervations | 302 | | | | | | | | |
| Standard deviation | 0.126 **T.HAUSMAN - 7086.989*** | | | | | | | | |

**(*) Coefficient statistically significant at 5%.**

This equation 3 estimated of real wages presents satisfactory results in terms of statistical significance of coefficients, the degree of adjustment and autocorrelation of errors. For the signs of the estimated coefficients that represent the respective elasticities, taking into account the expected by the economic theory, we confirm that, apart the gross value added, the price index and the nominal wages per employee, all coefficients have the expected signs.

**2.2. LINEARIZED AND REDUCED EQUATION OF REAL WAGES, WITH THE VARIABLES INDEPENDENT REGIONALLY DISAGGREGATED**

Following it is presented the equation of real wages reduced and in a linear form, but now with the independent variables disaggregated at regional level, in other words, considered only for the region being analyzed, and not for the whole of Portugal, as in the previous equation. Although this equation does not consider the effect of nearby regions of r in this region, aims to be a simulation to determine the effect of the regions in their real wages, that is:

$$\ln \omega_{rt} = f_0 + f_1 \ln Y_{rt} + f_2 \ln T_{rpt} + f_3 \ln G_{rt} + f_4 \ln \lambda_{rt} + f_5 \ln w_{rt} + f_6 \ln T_{prt} \qquad (4)$$

where:

- $\omega_{rt}$ is the real wage in the region r, for each of the manufacturing industries;
- Yrt is the gross value added of each of the manufacturing industries at the regional level;
- Grt is the price index at the regional level;
- $\lambda_{rt}$ is the number of workers in each industry, at regional level;
- Wrt is the nominal wage per employee in each of the manufacturing industries at regional level;
- Trpt is the flow of goods from each region to Portugal;
- Tprt is the flow of goods to each of the regions from Portugal.

Table 3 presents the results of estimating equation 4 where the independent variables are disaggregated at regional level. About the signs of the coefficients, it appears that these are the expected, given the theory, the same can not be said of the variable $\lambda_{rt}$ (number of employees). However, it is not surprising given the economic characteristics of regions like the Norte (many employees and low wages) and Alentejo (few employees and high salaries), two atypical cases precisely for opposite reasons. Analyzing the results in Tables 1, 2 and 3 we confirm the greater explanatory power of the variables when considered in aggregate at the national level.

**Table 3:** Estimation of the equation of real wages with the independent variables disaggregated at the regional level

$$\ln \omega_{rt} = f_0 + f_1 \ln Y_{rt} + f_2 \ln T_{rpt} + f_3 \ln G_{rt} + f_4 \ln \lambda_{rt} + f_5 \ln w_{rt} + f_6 \ln T_{prt},$$

| Variables | Const. | lnY$_{rt}$ | lnT$_{rpt}$ | lnG$_{rt}$ | ln $\lambda_{rt}$ | lnw$_{rt}$ | lnT$_{prt}$ | | |
|---|---|---|---|---|---|---|---|---|---|
| **Coefficients** | f$_0$* | f$_1$* | f$_2$* | f$_3$* | f$_4$* | f$_5$* | f$_6$* | R$^2$ | DW |
| **Random effects** Coefficients T-stat. L. signif. | 1.530 (3.355) (0.001) | 0.101 (4.147) (0.000) | 0.629 (4.625) (0.000) | -0.571 (-10.218) (0.000) | -0.151 (-5.364) (0.000) | 0.516 (13.357) (0.000) | -0.506 (-3.985) (0.000) | 0.670 | 1.858 |
| LSDV | | 0.098* (4.129) | 0.559* (4.449) | -0.624* (-11.380) | -0.155* (-6.130) | 0.619* (16.784) | -0.411* (-3.511) | 0.756 | 1.934 |
| Degrees of freedom | 295 - 289 | | | | | | | | |
| Number of obervations | 302 - 302 | | | | | | | | |
| Standard deviation | 0.155 - 0.165 **T.HAUSMAN - 72.843*** | | | | | | | | |

**(*) Coefficient statistically significant at 5%.**

### 2.3. ALTERNATIVE EQUATIONS TO THE EQUATIONS 3 AND 4

We also made two alternative estimates in order to test the existence of multicollinearity among the explanatory variables, considering all the variables by the weight of the work in every industry and every region in the total industry in this region for the equation 3 and the weight of work in every industry and every region in the national total of this industry for the equation 4, following procedures of (4)Hanson (1998). It is noted that the results are very similar to those previously presented to the estimates of equations 3 and 4, which allows us to verify the absence of statistics infractions.

### 2.4. EQUATION OF THE AGGLOMERATION

In the analysis of the Portuguese regional agglomeration process, using models of New Economic Geography in the linear form, we pretend to identify whether there are between Portuguese regions, or not, forces of concentration of economic activity and population in one or a few regions (centripetal forces). These forces of attraction to this theory, are the differences that arise in real wages, since locations with higher real wages, have better conditions to begin the process of agglomeration. Therefore, it pretends to analyze the factors that originate convergence or divergence in real wages between Portuguese regions. Thus, given the characteristics of these regions will be used as the dependent variable, the ratio of real wages in each region and the region's leading real wages in this case (Lisboa e Vale do Tejo), following procedures of Armstrong (1995) and Dewhurst and Mutis-Gaitan (1995). So, which contribute to the increase in this ratio is a force that works against clutter (centrifugal force) and vice versa.

Thus:

$$\ln\left(\frac{\omega_{rt}}{\omega_{lt}}\right) = a_0 + a_1 \ln Y_{nt} + a_2 \ln T_{rl} + a_3 \ln L_{nt} + a_4 \ln P_{rt} + a_5 \ln RL_{rmt} + a_6 \ln RL_{rgt} + a_7 RL_{rkt} + a_8 \ln RL_{rnt} \quad (5)$$

where:

- Ynt is the national gross value added of each of the manufacturing industries considered in the database used;
- Trl is the flow of goods from each region to Lisboa e Vale do Tejo, representing the transportation costs;
- Lnt is the number of employees in manufacturing at the national level;
- Prt is the regional productivity (ratio of regional gross value added in manufacturing and the regional number of employees employed in this activity);
- RLrmt is the ratio between the total number of employees in regional manufacturing and the regional number of employees, in each manufacturing (agglomeration forces represent inter-industry, at regional level);
- RLrgt is the ratio between the number of regional employees in each manufacturing and regional total in all activities (represent agglomeration forces intra-industry, at regional level);
- RLrkt is the ratio between the number of regional employees in each manufacturing, and regional area (representing forces of agglomeration related to the size of the region);

- RLrnt is the ratio between the number of regional employees, in each of the manufacturing industries, and the national total in each industry (agglomeration forces represent inter-regions in each of the manufacturing industries considered).

The index r (1,..., 5) represents the respective region, t is the time period (8 years), n the entire national territory, k the area (km2), l the region Lisboa e Vale do Tejo, g all sectors and m manufacturing activity (9 industries).

The results of the estimations made regarding equation 5 are shown in Tables 4 and 5. Two different estimates were made, one without the variable productivity (whose results are presented in Table 4) and one with this variable (Table 5).

**Table 4:** Estimation of the agglomeration equation without the productivity

$$\ln\left(\frac{\omega_{rt}}{\omega_{lt}}\right) = a_0 + a_1 \ln Y_{nt} + a_2 \ln T_{rl} + a_3 \ln L_{nt} + a_4 \ln RL_{rmt} + a_5 \ln RL_{rgt} + a_6 RL_{rkt} + a_7 \ln RL_{rnt}$$

| Variab. | Constant | lnY$_{nt}$ | lnT$_{rl}$ | lnL$_{nt}$ | lnRL$_{rmt}$ | lnRL$_{rgt}$ | lnRL$_{rkt}$ | lnRL$_{rnt}$ | | |
|---|---|---|---|---|---|---|---|---|---|---|
| Coef. | a$_0$ | a$_1$ | a$_2$ | a$_3$ | a$_4$ | a$_5$ | a$_6$ | a$_7$ | R$^2$ | DW |
| **Random ef.** V.Coef. T-stat. L. sign. | -3.991 (-3.317) (0.001) | -0.040 (-1.353) (0.177) | 0.012 (1.469) (0.143) | 0.390 (4.046) (0.000) | -0.413 (-4.799) (0.000) | -0.507 (-4.122) (0.000) | -0.228 (-4.333) (0.000) | 0.368 (4.249) (0.000) | 0.253 | 1.474 |
| Degrees of freedom | 293 | | | | | | | | | |
| Number of obervations | 302 | | | | | | | | | |
| Standard deviation | 0.126 **T.HAUSMAN - 1.870** | | | | | | | | | |

**(\*) Coefficient statistically significant at 5%.**
**(\*\*) Coefficient statistically significant at 10%.**

**Table 5:** Estimation of the agglomeration equation with the productivity

$$\ln\left(\frac{\omega_{rt}}{\omega_{lt}}\right) = a_0 + a_1 \ln Y_{nt} + a_2 \ln T_{rl} + a_3 \ln L_{nt} + a_4 \ln P_{rt} + a_5 \ln RL_{rmt} + a_6 \ln RL_{rgt} + a_7 RL_{rkt} + a_8 \ln RL_{rnt}$$

| Variab. | Constant | lnY$_{nt}$ | lnT$_{rl}$ | lnL$_{nt}$ | lnP$_{rt}$ | lnRL$_{rmt}$ | lnRL$_{rgt}$ | lnRL$_{rkt}$ | lnRL$_{rnt}$ | | |
|---|---|---|---|---|---|---|---|---|---|---|---|
| Coef. | a$_0$* | a$_1$* | a$_2$* | a$_3$* | a$_4$* | a$_5$* | a$_6$* | a$_7$* | a$_8$* | R$^2$ | DW |
| **Random eff.** V.Coef. T-stat. L. sign. | -3.053 (-2.991) (0.003) | -0.240 (-7.182) (0.000) | 0.015 (2.026) (0.044) | 0.486 (5.934) (0.000) | 0.218 (8.850) (0.000) | -0.266 (-3.494) (0.001) | -0.333 (-3.102) (0.002) | -0.141 (-3.067) (0.002) | 0.230 (3.026) (0.003) | 0.455 | 1.516 |
| LSDV | -0.307* (-9.259) | -0.033* (-4.821) | 0.330* (5.701) | 0.256* (8.874) | -0.049 (-0.972) | 0.011 (0.169) | -0.027 (-0.968) | 0.006 (0.137) | | 0.649 | 1.504 |
| Degrees of freedom | 292 - 285 | | | | | | | | | | |
| Number of obervations | 302 - 302 | | | | | | | | | | |
| Standard deviation | 0.116 - 0.136 **T.HAUSMAN - 33.578\*** | | | | | | | | | | |

**(\*) Coefficient statistically significant at 5%.**

Comparing the values of two tables is confirmed again the importance of productivity (Prt) in explaining the wage differences. On the other hand improves the statistical significance of coefficients and the degree of explanation.

### 3. SOME FINAL CONCLUSIONS

With the above analysis, it appears that the explanatory power of the independent variables considered in models of New Economic Geography, is more reasonable, even when these variables are considered in their original form, in other words, in the aggregate form for all locations with strong business with that we are considering (in the case studied, aggregated at national level to mainland Portugal). However, the agglomeration process of the Portuguese regions, analyzing the set of coefficients of the estimations, in Lisboa e Vale do Tejo is not impressive, but when we look at the data this region has a greater potential of attractiveness of the population and economic activity. This is because that's where real wages are more uniform across different industries and higher than in other regions. However, the estimation results reflect some strange situations, in the face of the theory, namely the fact the Norte has the highest value of employees in manufacturing, the highest gross value added in this industry, but has the lowest real wages, explained possibly by the great weight of the textile industry in this region. The same we verify, but precisely in the contrary to the Alentejo. Perhaps, a finer spatial unit could help to explain these strange situations, but the lack of data for the NUTS III prevents this analysis. Anyway, the direct effect of considering large spatial units is reduced (as can be seen in Table 5 with the value obtained for the variable $RL_{rkt}$, or -0141). Despite some inconsistencies found in the face of the theory, it was possible to identify a set of centripetal forces (forces that favor the agglomeration) and a set of centrifugal forces (forces that work against agglomeration).

On the other hand, given the existence of "backward and forward" linkages and agglomeration economies, represented in the variables $RI_{rmt}$ and $RL_{rgt}$, we can affirm the existence of growing scale economies in the Portuguese manufacturing industry during the period considered. This taking into account the mentioned by (5)Marshall (1920) which in modern terminology argued that increasing returns to scale occur in industry, in the face of "spillover" effects, advantages of market expertise and "backward" and "forward" linkages associated with large local markets. Therefore, the trend during this period was for the regional divergence in Portugal, considering what referred by Hanson (1998), in other words, "The interaction of scale economies and transport costs creates a centripetal force, to use Krugman's language, that causes firms to agglomerate in industry centers".

It should be noted also that different estimates were made without the productivity variable and with this variable in order to be analyzed the importance of this variable in explaining the phenomenon of agglomeration. It seems important to carry out this analysis, because despite the economic theory consider the wages that can be explained by productivity, the new economic geography ignores it, at least explicitly, in their models, for reasons already mentioned widely, particular those related to the need to make the models tractable.

Finally, is important to refer the importance of the transportation costs in explaining the spatial issues, reinforced by the fact that the estimates made with the seven NUTS II Portugal (including Madeira and Açores) present values much worse than when considering only the five NUTS II. What makes sense, since the real wage developments do not follow the increase in transport costs from the continent for these two Portuguese islands.

## 4. REFERENECES